\documentclass[%
 reprint,
 amsmath,amssymb,
 aps,
]{revtex4-2}

\usepackage{graphicx}
\usepackage{dcolumn}
\usepackage{bm}

\usepackage{graphicx}

\usepackage{float}
\usepackage{amsmath} 
\usepackage{booktabs}  
\usepackage{longtable} 
\usepackage{pgfplotstable} 
\usepackage{hyperref}

\usepackage{epstopdf}  
\usepackage{amssymb}
\usepackage{array}
\usepackage{pifont}

\usepackage{listings}
\usepackage{changepage}

\begin{document}

\preprint{APS/123-QED}

\title{Large Language Models for Superconductor Discovery}


\author{Suman Itani$^{1,\dagger}$}
\author{Yibo Zhang$^{1,\dagger}$}
\author{Ranjit Itani$^{2}$}
\author{Jiadong Zang$^{1}$}


\affiliation{$^{1}$Department of Physics and Astronomy, University of New Hampshire, 9 Library Way, Durham, NH 03824, USA}

\affiliation{$^{2}$Department of Electrical and Computer Engineering, University of New Hampshire, 33 Academic Way, Durham, NH 03824, USA}


\thanks{$^{\dagger}$Corresponding author: suman.itani@unh.edu,\\ yibo.zhang@unh.edu}

\date{\today}

\begin{abstract}
Large language models (LLMs) offer new opportunities for automated data extraction and property prediction across materials science, yet their use in superconductivity research remains limited. Here we construct a large experimental database of 78,203 records, covering 19,058 unique compositions, extracted from scientific literature using LLM driven workflow. Each entry includes chemical composition, critical temperature, measurement pressure, structural descriptors, and critical fields. We fine-tune several open-source LLMs for three tasks: (i) classifying superconductors vs. non-superconductors, (ii) predicting the superconducting transition temperature directly from composition or structure-informed inputs, and (iii) inverse design of candidate compositions conditioned on target T$_c$. The fine-tuned LLMs achieve performance comparable to traditional feature-based models---and in some cases exceed them---while substantially outperforming their base versions and capturing meaningful chemical and structural trends. The inverse-design model generates chemically plausible compositions, including 28\% novel candidates not seen in training. Finally, applying the trained predictors to GNoME database identifies unreported materials with predicted $T_c > 10 K$. 
Although unverified, these candidates illustrate how integrating LLM-driven workflow can enable scalable hypothesis generation for superconductivity discovery. 

\end{abstract}

\maketitle


\section{\label{sec:level1}INTRODUCTION}

Superconductivity is a quantum phenomenon in which a material exhibits zero electrical resistance and expels magnetic fields below a critical temperature (\(T_c\)). The relentless pursuit of superconductors with higher transition temperatures is driven by their transformative potential in energy-efficient power transmission, high-field magnets, magnetic levitation, and quantum technologies \cite{hassenzahl2004electric,gambetta2017building}. Realizing these applications requires materials that maintain superconductivity at high~$T_c$ under practical and affordable cooling conditions. To date, only a few material families---primarily cuprates \cite{bednorz1986possible} and iron-based superconductors at ambient pressure \cite{kamihara2008iron}, and hydrogen-rich hydrides at extreme pressures exceeding 150 GPa \cite{drozdov2015conventional}---have demonstrated \(T_c\) values above 100~K. Achieving room-temperature superconductivity at ambient pressure remains one of the most compelling and unresolved challenges in condensed matter physics and materials science.

Identifying new superconductors remains challenging due to the absence of a universal predictive theory. The Bardeen–Cooper–Schrieffer (BCS) theory~\cite{bardeen1957theory,bardeen1957microscopic} successfully explains phonon-mediated superconductivity, but many high-\(T_c\) materials exhibit strong electronic correlations or unconventional pairing mechanisms that lie beyond the scope of BCS theory. Solving the full many-body Schrödinger equation for realistic materials is computationally infeasible, while density functional theory often underestimates key interactions in strongly correlated systems. As a result, superconductor discovery has historically relied on empirical rules and trial-and-error synthesis, with only a small fraction of chemically accessible compounds exhibiting superconductivity~\cite{matthias1957chapter}. Given the vast compositional and structural design space, purely heuristic approaches alone are no longer sufficient. Data-driven techniques offer a promising complementary pathway by uncovering patterns embedded in known superconductors that may accelerate the identification of new candidates.\\
The use of patterns in data to drive scientific discovery has a long history; Mendeleev’s 1869 periodic table is a landmark example, and modern machine learning extends this principle to high-dimensional materials datasets. Building on this perspective, the availability of high-throughput computational resources and curated databases such as SuperCon has enabled the use of machine learning to predict superconducting properties directly from chemical composition~\cite{materials-a}.
Early successes in this domain demonstrated the feasibility of both classification and regression tasks. Stanev et al.~\cite{stanev2018machine} applied a random forest classifier to a few thousand compounds and showed that machine learning can reliably distinguish superconductors from non-superconductors. Hamidieh~\cite{hamidieh2018data} developed a gradient-boosting model trained on a few thousand known superconductors using elemental descriptors to predict critical temperatures. These efforts marked a shift from manual heuristics to data-driven inference.
Subsequent works introduced deep learning architectures tailored to composition-based representations. Konno et al.~\cite{konno2021deep} encoded chemical formulas onto a periodic-table grid and trained convolutional neural networks to recognize superconducting patterns. Pereti et al.~\cite{pereti2023individual} employed a DeepSet framework to model unordered element sets and successfully screened minerals for telluride-based superconductors, validating their predictions experimentally. Other approaches have incorporated convolutional and recurrent layers~\cite{li2020critical}, or trained deep neural networks directly on electronic band structures to infer phase diagrams~\cite{li2025deep}. To improve robustness and uncertainty quantification, ensemble methods such as Optuna-optimized stacking~\cite{yu2023prediction}, variational Bayesian networks, and Monte Carlo dropout have also been explored~\cite{le2020critical}.
Beyond these examples, diverse machine learning models have been applied to superconductivity classification and \(T_c\) prediction~\cite{gashmard2024predicting,kaplan2025deep,bai2024unveiling,jiang2025machine,jung2024machine,xie2022machine,lesser2025learning,gu2024bond,nieto2024predicting,pogue2023closed,
chen2025pscg}. These efforts reflect the growing impact of data-driven methods in accelerating superconductor discovery.

Although these data-driven models demonstrate that composition alone encodes useful information, they also reveal significant limitations. Most training sets contain only a few thousand entries and rely on manually engineered descriptors---Magpie features~\cite{ward2016general}, periodic-table images, or other physically motivated quantities---that may bias the learned relationships and limit generalization. Structural information is critical for understanding superconductivity, yet the SuperCon database---the most widely used resource---contains about 16{,}400 compounds but provides only chemical compositions and critical temperatures; both pressure data and full crystal structures are absent~\cite{materials-a}.  Models trained solely on composition cannot distinguish structural polymorphs or capture the influence of lattice geometry on \(T_c\). Efforts to incorporate structure, such as 3DSC \cite{sommer20233dsc}, S2SNet \cite{liu2023s2snet} dataset that matched SuperCon compositions with approximate structures from the Materials Project \cite{jain2013commentary}, ICSD \cite{bergerhoff1983inorganic, zagorac2019recent}, improve predictive performance but reduce the number of usable entries because structural data exist for only a subset of compounds. Another limitation is the limited availability of verified non-superconductors. While some negative entries are present, they are far fewer than positive cases, hindering the development of balanced and reliable classifiers.

A major bottleneck in building comprehensive and accurate datasets is that most experimental data reside in unstructured scientific publications and are not readily machine‑readable.  Manually reading papers to extract structured informatio is labour‑intensive.  Rule‑based natural‑language‑processing systems, such as ChemDataExtractor \cite{swain2016chemdataextractor, court2020magnetic}, can automate some extraction tasks but require extensive domain expertise and often struggle with complex sentence structures.  Recently, large language models (LLMs) trained on massive corpora have shown remarkable capabilities in information extraction \cite{zhang2024gptarticleextractor,zhang2024fine}. Carefully engineered prompts already enable GPT‑4o to parse important structured information for magnetic and thermoelectric compounds with high accuracy\cite{itani2025northeast, itani2025large}.

Beyond extraction, fine‑tuned versions of GPT‑3 and related models can answer chemistry questions, perform property prediction and classify materials phases using only modest amounts of task‑specific data \cite{jablonka2024leveraging}. Also, language models have been adapted for generative materials design.  AtomGPT encodes chemical formulas and crystal structures as text and fine‑tunes transformer architectures to predict different materials properties \cite{choudhary2024atomgpt}.  DiffractGPT uses a generative transformer to reconstruct atomic structures directly from X‑ray diffraction patterns \cite{choudhary2025diffractgpt}.  Fine‑tuned language models have also been shown to generate chemically valid and stable inorganic compounds with higher success rates than diffusion models \cite{cao2025crystalformer, gruver2024fine,antunes2024crystal}. 
Collectively, these studies suggest that, when properly adapted, large language models can function as general-purpose engines for property prediction, structure inference, and even inverse design. However, the application of LLMs to superconductivity---spanning comprehensive data extraction as well as forward and inverse property prediction---remains in its early stages and has thus far been limited to relatively small, curated datasets.

\begin{figure*}
\includegraphics[width=\textwidth]{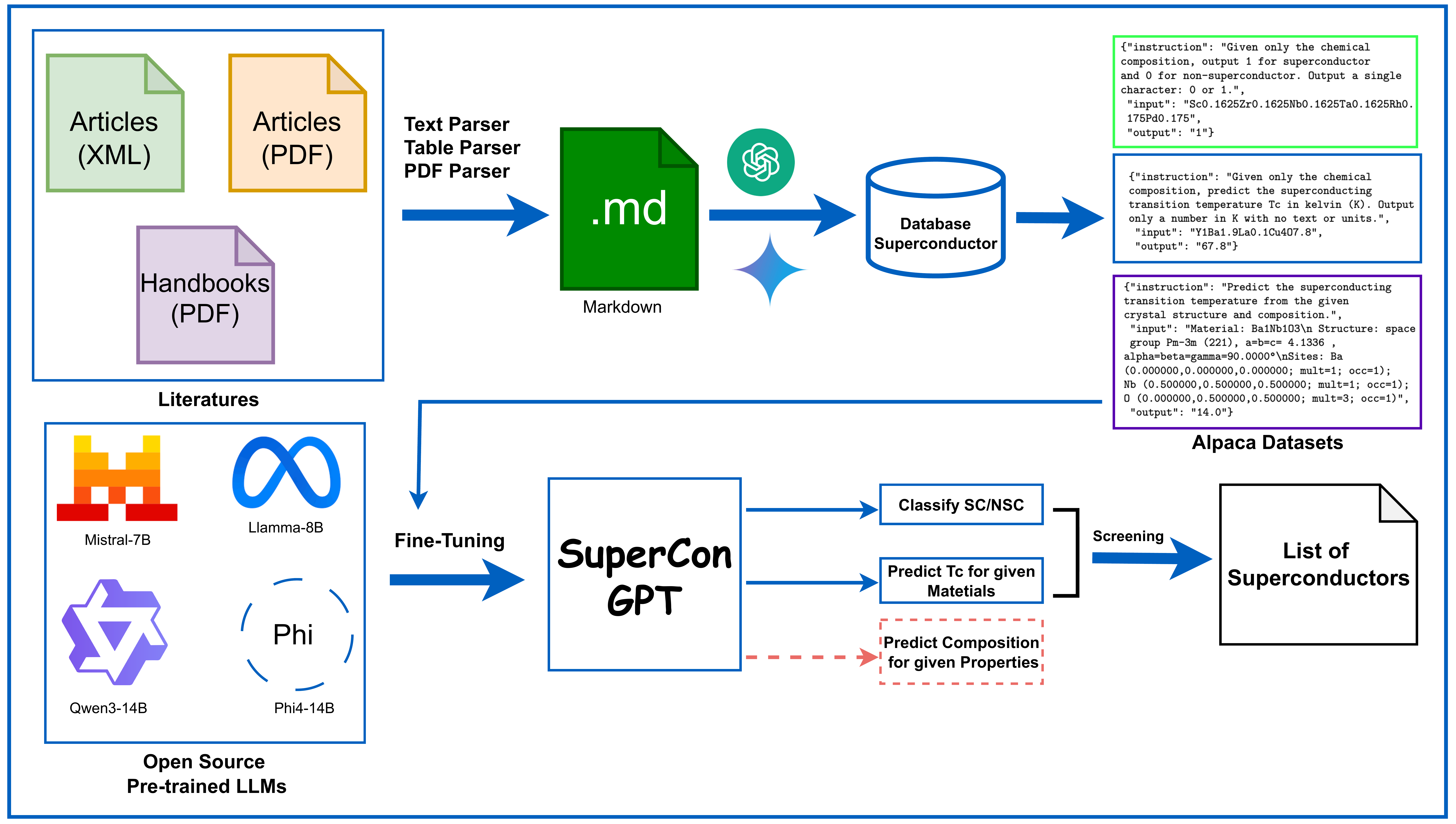}
\caption{\label{fig:workflow2}
\textbf{End-to-end LLM-driven workflow.}
Scientific articles and handbooks are parsed into markdown files, which are processed by an LLM to extract structured data. The resulting database is used to generate Alpaca-format instruction datasets for fine-tuning open-source LLMs (Mistral-7B, Llama~3.1-8B, Qwen3-14B, and Phi-14B). The fine-tuned models perform SC/NSC classification, $T_c$ regression, and inverse design, enabling large-scale identification of candidate superconductors.}
\end{figure*}

In this work, we use large language models across the entire data–extraction-to-property-prediction pipeline, as shown in Figure \ref{fig:workflow2}. First, we extend our previously developed automated text-mining framework~\cite{itani2025northeast} with superconductivity-specific prompts to extract structured information directly from the literature. The system identifies relevant papers, parses chemical compositions, records reported transition temperatures along with applied pressures, and captures structural descriptors such as crystal system, space group, and lattice parameters when available. Using this workflow, we constructed a dataset of 78,204 entries, covering about 19,058 unique compositions---including many not present in existing databases. The dataset contains both superconducting and non-superconducting reports and preserves multiple \(T_c\) values for compositions measured under different pressures or doping conditions.

Second, we fine-tuned open-source transformer models (Mistral-7B, Llama-3.1-8B, Qwen3-14B, Qwen3-2507-4B, Phi4-14B) for two supervised tasks: (i) classifying whether a material is superconducting and (ii) predicting its critical temperature. Because these models operate directly on textual representations of chemical formulas and  structural metadata, they bypass the need for hand-engineered descriptors. We systematically compared the performance of all fine tuned LLMs across multiple input formats and benchmarked them against feature-based models such as random forests, XGBoost, and fully connected neural networks. The best transformer classifier (Qwen3-14B) achieved an accuracy of 91.1\%, essentially matching the performance of the best feature-based baseline (91.0\%), while showing stronger generalization to compositions outside the training distribution. For \(T_c\) regression, the best-performing fine-tuned LLMs reached \(R^2 \approx 0.84\), significantly outperforming their base counterparts and achieving accuracy comparable to feature-based models. We also fine-tuned Qwen3-14B for inverse design, allowing it to generate compositions and space groups conditioned on a target \(T_c\). In our tests, the model accurately reproduced about 73\% of the compositions seen in the training set and generated 27\% novel compositions not present in the training data.

Finally, we applied the trained LLM predictors to GeNoME database to identify potential new superconductors. These models identified -- previously unreported materials as potential superconductor candidates. While these predictions have not yet been experimentally validated, they illustrate how combining LLM-based literature extraction with LLM-based prediction provides a scalable framework for accelerating the discovery of superconducting materials.

\section{\label{sec:level1}Methodology}

\subsection{\label{sec:level2}Automated Database Compilation Using LLMs}

We constructed the superconductivity database using an automated pipeline driven by large language models (LLMs), following and extending our earlier approaches~\cite{zhang2024gptarticleextractor,itani2025northeast} . The workflow, shown in Figure~\ref{fig:workflow2}, begins by identifying relevant scientific papers from sources such as the American Physical Society (APS) and Elsevier. Using keyword searches including ``superconductor,'' ``critical temperature,'' ``superconducting materials,'' and ``non-superconductor,'' we collected approximately 90{,}000 unique DOIs.

Many of these articles were available directly in XML format through publisher APIs. These XML files were processed with dedicated parsers to generate structured Markdown documents. However, older APS articles and several handbooks were available only as scanned PDFs containing valuable historical data. These PDF documents were converted to Markdown using specialized text, table, and PDF parsers, with assistance from the Google Gemini LLM.

All Markdown documents were then passed through a set of carefully engineered prompts for advanced LLMs (e.g., GPT-4o, Gemini). The models extracted structured information---including compositions, reported transition temperatures, applied pressures, and available structural metadata---and returned the results in JSON format. These JSON outputs were aggregated  to produce the final comprehensive superconductivity database.

\subsection{\label{sec:level2}Fine-tuning Large Language Models}

To adapt foundation models to domain-specific superconductivity prediction tasks, we fine-tuned Mistral-7B~\cite{jiang2023mistral7b}, LLaMA 3.1-8B~\cite{grattafiori2024llama}, Qwen3-14B~\cite{yang2025qwen3}, Qwen3-2507-4B, and Phi4-14B~\cite{abdin2024phi} using a parameter-efficient fine-tuning (PEFT) approach based on Low-Rank Adaptation (LoRA)~\cite{hu2022lora}, implemented through the Unsloth package \cite{unsloth}. These models were selected because they provide strong general-purpose performance while maintaining compact model sizes, making them feasible to fine-tune on our available GPU resources (a single 24~GB GPU).

In our implementation, the pretrained model was loaded in 4-bit quantization for memory efficiency and wrapped with LoRA adapters targeting both the self-attention projections (\verb|q_proj|, \verb|k_proj|, \verb|v_proj|, \verb|o_proj|) and the feed-forward projections (\verb|gate_proj|, \verb|up_proj|, \verb|down_proj|). The LoRA rank and scaling factors were set to $r = 16$ and $\alpha = 16$, respectively, and dropout was disabled for determinism. Gradient checkpointing was enabled to minimize activation memory usage during backpropagation.  

For a linear transformation with weight matrix $W \in \mathbb{R}^{d_{\text{out}} \times d_{\text{in}}}$, LoRA augments it as $W' = W + \Delta W$, where $\Delta W = BA$, with $A \in \mathbb{R}^{r \times d_{\text{in}}}$ and $B \in \mathbb{R}^{d_{\text{out}} \times r}$, and $r \ll \min(d_{\text{in}}, d_{\text{out}})$. During training, only the low-rank matrices $A$ and $B$ are updated, leaving the pretrained weights frozen. This approach allows the model to specialize efficiently for materials-science data while preserving the general linguistic and reasoning knowledge learned during large-scale pretraining.

\subsubsection{Dataset preparation and instruction templates.}

All models were fine-tuned using a supervised instruction-tuning paradigm, with training samples formatted in the Alpaca style~\cite{alpaca}. Each record contains three fields---\verb|instruction|, \verb|input|, and \verb|output|---which are concatenated into a single training sequence and terminated with an end-of-sequence token (\verb|eos_token|). The resulting text string is stored under a single field, \verb|text|, for use with the supervised fine-tuning trainer. This unified format is compatible with multiple task types---including binary classification, numerical regression, and inverse design---while each model is fine-tuned separately on the dataset corresponding to its specific task.

\medskip

\paragraph{Binary classification format.}  
For the superconductor/non-superconductor classification task, each input consisted only of the chemical composition, and the model was required to output \verb|0| or \verb|1|:

\begin{quote}
\ttfamily\small\raggedright
{"instruction": "Given only the chemical composition, output 1 for superconductor
and 0 for non-superconductor. Output a single character: 0 or 1.",
 "input": 
"Sc0.1625Zr0.1625Nb0.1625Ta0.1625Rh0.175Pd0.175",
 "output": "1"}
\end{quote}

We restrict the output vocabulary to numeric tokens (\verb|0| or \verb|1|) rather than textual labels (e.g., ``superconductor''), which avoids token overlap and simplifies the cross-entropy loss.

\medskip

\paragraph{Composition only-based regression format.}  
For predicting the superconducting transition temperature (\(T_c\)) from composition alone, we used:

\begin{quote}
\ttfamily\small\raggedright
{"instruction": "Given only the chemical composition, predict the superconducting
transition temperature Tc in kelvin (K). Output only a number in K with no text or units.",
 "input":
"Y1Ba1.9La0.1Cu4O7.8",
 "output": "67.8"}
\end{quote}

During training, the model minimizes the token-level cross-entropy loss; at evaluation, the output string is parsed as a floating-point value to compute MAE, RMSE, and \(R^2\).

\medskip

\paragraph{Regression with crystal system and space group.}  
To incorporate structural metadata, we designed a second regression dataset that included crystal system, space group, and pressure (when available):

\begin{quote}
\ttfamily\small\raggedright
{"instruction": "Given a material's crystal system, space group, and chemical composition,
predict its superconducting transition temperature (Tc).",
 "input":
"Material: Eu1Fe1.62Co0.38As2
Crystal system: Tetragonal
Space group: I4/mmm
Pressure: ambient",
 "output": "5.15 K"}
\end{quote}

\medskip

\paragraph{Regression with full CIF information.}  
We additionally constructed a structure-informed dataset using complete crystallographic information files (CIFs). The corresponding template was:

\begin{quote}
\ttfamily\small\raggedright
{"instruction": "Predict the superconducting transition temperature from the given
crystal structure and composition.",
 "input":
"Material: Ba1Nb1O3
Structure: space group Pm-3m (221), a=b=c=4.1336 Å, alpha=beta=gamma=90.0000°
Sites: Ba (0.000000,0.000000,0.000000; mult=1; occ=1);
      Nb (0.500000,0.500000,0.500000; mult=1; occ=1);
      O  (0.000000,0.500000,0.500000; mult=3; occ=1)",
 "output": "14.0"}
\end{quote}

Encoding lattice parameters, atomic positions, and symmetry information in serialized text enables the transformer to attend jointly to structural and compositional descriptors through its self-attention mechanism.

\medskip

\paragraph{Inverse design format.}  
Finally, we prepared an inverse design dataset in which the model is asked to propose a superconducting composition along space group, conditioned on a target \(T_c\) and, optionally, a crystal system:

\begin{quote}
\ttfamily\small\raggedright
{"instruction": "Given a target Tc and crystal system, propose a likely superconducting composition.",
 "input":
"Target Tc: 92 K
Crystal system: Orthorhombic",
 "output": "Y0.997Fe0.003Ba2Cu3O7 (space group Pmmm, under ambient pressure); reported Tc = 92.3 K."}
\end{quote}

This dataset allows the model to learn the reverse mapping from desired properties to plausible compositions and structural motifs.

\subsubsection{Training objective and optimization}
Fine-tuning was performed using the Unsloth framework~\cite{unsloth} together with the \verb|SFTTrainer| module from the Transformers Reinforcement Learning (TRL) library~\cite{vonwerra2022trl}. All models were trained for four epochs using AdamW (8-bit) with a weight decay of 0.01, a learning rate of $2\times 10^{-4}$, and a linear warm-up schedule. Each GPU processed a batch size of 2 with gradient accumulation of 4, giving an effective batch size of 8. Mixed-precision training (fp16/bf16) was employed to reduce memory usage and improve throughput.

During fine-tuning, the model receives tokenized instruction--input--output sequences. Each token is embedded and propagated through a stack of transformer blocks, where information is integrated through multi-head self-attention and position-wise feed-forward layers. For a single attention head, the scaled dot-product attention is computed as
\begin{equation}
\mathrm{Attention}(Q,K,V)
= \mathrm{softmax}\!\left(\frac{QK^{\top}}{\sqrt{d_k}}\right)V,
\label{eq:attention}
\end{equation}
where $Q$, $K$, and $V$ denote the query, key, and value matrices and $d_k$ is the key dimensionality \cite{vaswani2017attention}. Multi-head attention concatenates multiple such heads, enabling the model to attend to different chemical and structural patterns across the serialized sequence.

Supervised fine-tuning optimizes the causal language-modeling cross-entropy loss over the output portion of each training sequence.
Given a target token sequence $y = (y_1,\ldots,y_T)$ and model
distribution $p_\theta$, the loss is
\begin{equation}
\mathcal{L}_{\mathrm{CE}}
= -\sum_{t=1}^{T} \log p_\theta\!\left(y_t \mid y_{<t}, x\right),
\label{eq:cross_entropy}
\end{equation}
where $x$ denotes the concatenated instruction and input.
During training, only the response tokens contribute to the loss,
ensuring that the model learns the desired task mapping
(classification, regression, or inverse design).

This instruction-driven formulation enables a unified training setup across all kind of datasets. Parameter-efficient fine-tuning via LoRA adapters allows the transformer to specialize in superconductivity prediction while keeping the base model weights frozen, resulting in domain adaptation with minimal computational overhead.

\subsection{\label{sec:level2}Feature-based Machine Learning Models}

To benchmark the fine-tuned language models, we trained a set of traditional feature-based models for both superconductivity classification and $T_c$ regression. These models train on manually engineered numerical descriptors.For each composition, we first built an elemental feature vector in which each element is assigned a normalized fractional proportion in the range 0–1 (0 for absent elements, 1 for a pure element). Additional scalar descriptors (e.g., average atomic weight, average atomic number, average electronegativity, and related compositional statistics) were computed using \texttt{pymatgen} library~\cite{ong2013python,ward2016general}. 


For the SC/NSC classification task, we trained Random Forest and XGBoost classifiers on the engineered feature set. The data were split into stratified training, validation, and test subsets (overall 80:20 train–test ratio), preserving the class distribution. Hyperparameters were tuned on the validation set, and the final models were evaluated on the held-out test set in terms of accuracy, precision, recall, F$_1$-score, and confusion matrices.

For $T_c$ prediction, we used the same feature construction pipeline. We again applied a stratified train–test split, using quantile bins in $\log_{10}(T_c + 1)$ to balance different temperature ranges. As baseline regressors, we trained (i) an XGBoost ensemble and (ii) a fully connected feedforward neural network ensemble (iii) random forest. In all cases, multiple models were trained on stratified, down-sampled subsets of the training data, and predictions on the test set were obtained by averaging across ensemble members. Performance was quantified using the coefficient of determination ($R^2$), mean absolute error (MAE), and root mean square error (RMSE), and was used as a reference point for assessing the gains achieved by the fine-tuned LLMs.

\section{\label{sec:level1}Results and Discussion}
\subsection{\label{sec:level2}Construction of the Experimental Superconductor Database}

We developed an experimental superconductor database using the workflow illustrated in Figure~\ref{fig:workflow2}. A key strength of our approach is that, in addition to compiling superconductors and their associated properties, we also incorporated experimentally verified non-superconductors. This inclusion is critical for training reliable machine learning classifiers capable of distinguishing superconductors from non-superconductors.

The final database contains 78,203 entries, aggregated from sources such as the American Physical Society (APS), Elsevier journals, and domain handbooks on superconductivity. These records correspond to 19,058 unique chemical compositions. For each entry, we extracted the material name, chemical formula, material class (e.g., bulk or thin film), and material type (superconductor or non-superconductor). critical temperature ($T_c$) values were recorded together with the applied measurement pressure, and, where available, both lower and upper critical magnetic fields were included. We also collected structural descriptors such as crystal system, lattice type, lattice parameters, and space group. Each record explicitly indicates whether the information originates from experimental measurements or theoretical calculations. A complete list of extracted fields is provided in Table~\ref{table:features}.

The reliability of this workflow is supported by our previous construction of the Northeast Materials Database (NEMAD), which uses the same LLM-based extraction methodology and was validated through expert review\cite{itani2025northeast}. This prior validation demonstrates the robustness of the approach and provides confidence in the accuracy of the superconductivity database generated here.

A comparison with the widely used SuperCon database shows that our dataset serves as a complementary resource that expands the available experimental information on superconducting materials. Whereas SuperCon primarily reports chemical composition and critical temperature, our database also records structural descriptors, applied pressure, critical fields, and material-class annotations. Among the $\sim$19,058 unique compositions in our collection, approximately 14,000 do not appear in SuperCon. These additional compositions and property types provide broader coverage of the superconducting materials landscape and offer a valuable supplementary resource for both superconductivity research and data-driven materials discovery.

\begin{figure*}[ht]
\includegraphics[width=\textwidth]{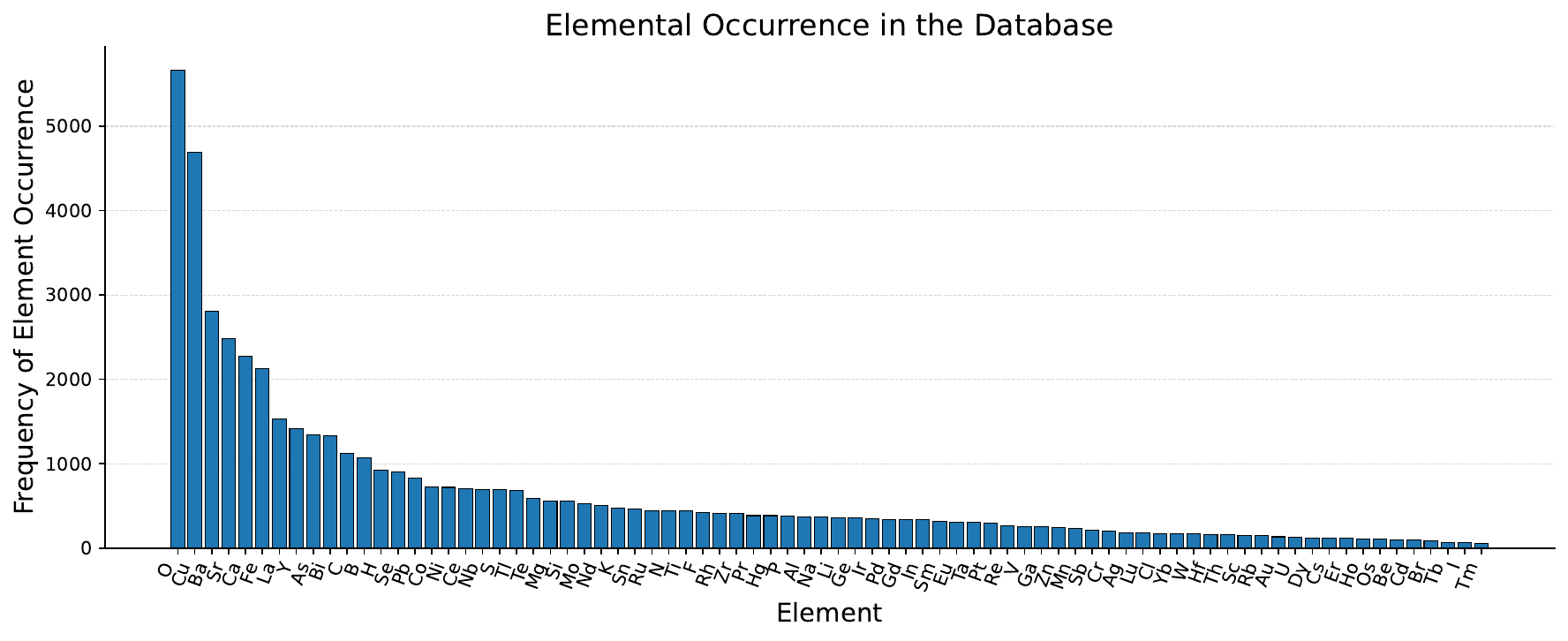}
\caption{\label{fig:occurance}
\textbf{Elemental occurrence frequency in the superconductor database.}
The bar chart shows the number of distinct compounds in which each element appears, based on the curated superconductor dataset. All elements present in the database are included, with oxygen and copper showing the highest occurrence, reflecting their dominant role in known superconducting materials.}
\end{figure*}

The frequency distribution of chemical elements in the database (Figure~\ref{fig:occurance}) shows clear trends consistent with the chemistry of known superconductors. Oxygen appears in the largest number of compounds, followed by copper, barium, strontium  and other rare-earth or transition-metal elements frequently found in cuprate and perovskite-derived superconductors. This dominance reflects both the historical research emphasis on oxide-based superconductors and the structural tunability of these materials. For clarity, some of these very low-frequency elements are not shown in the figure. Overall, the long-tailed distribution highlights the chemical bias of the literature: a relatively small subset of elements accounts for the majority of experimentally studied superconducting compositions.

\begin{figure}[ht]
\includegraphics[width=0.48\textwidth]{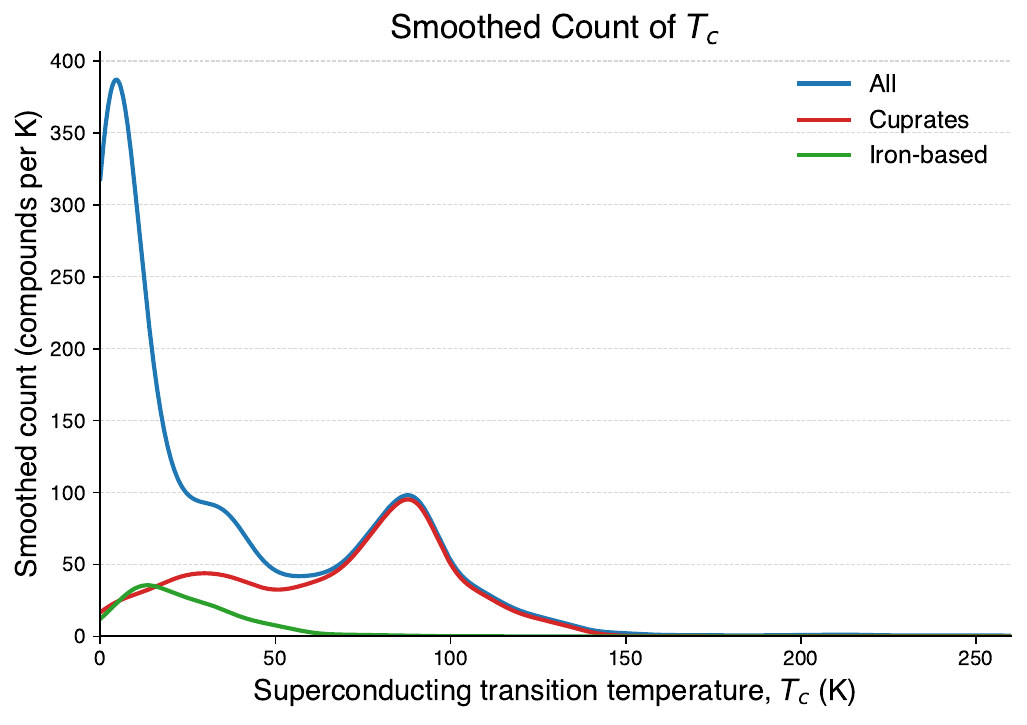}
\caption{\label{fig:tc_distribution}
\textbf{Smoothed distribution of superconducting transition temperatures ($T_c$) in the compiled database.}
Kernel-smoothed counts are shown for all superconductors (blue), cuprate families (red), and iron-based compounds (green). 
The distribution exhibits a strong peak at low temperatures and a secondary maximum near 90--100~K associated with cuprate high-$T_c$ materials.}
\end{figure}

The distribution of superconducting transition temperatures (Figure~\ref{fig:tc_distribution}) further illustrates this heterogeneity. A strong peak is observed below 20~K, corresponding to the large number of conventional BCS superconductors. A second prominent peak emerges near 90--100~K, dominated by cuprate compounds, while the iron pnictide superconductors form a smaller but distinct peak around 30--40~K. These multimodal trends reflect the fundamental grouping of the superconducting families and show that the dataset captures the full diversity of experimentally explored superconductors. The smoothed density curves also reveal the scarcity of materials with $T_c$ above $\sim$150~K, underscoring the continued difficulty of discovering new high-$T_c$ materials.


\begin{table}[ht]
\caption{\label{table:features}
\textbf{Summary of database features and their types.}}
\begin{ruledtabular}
\begin{tabular}{lcc}
Feature & Type & Unit \\
\hline
Material Name         & String  & -- \\
Chemical Composition  & String  & -- \\
Material Class        & String  & -- \\
Material Type         & String  & -- \\
Superconductor Type   & String  & -- \\
$T_\mathrm{c}$         & Numeric & K  \\
Pressure              & Numeric & Pa \\
Lower Critical Field  & Numeric & T  \\
Upper Critical Field  & Numeric & T  \\
Crystal Structure     & String  & -- \\
Lattice Structure     & String  & -- \\
Lattice Parameters    & Numeric & \AA \\
Space Group           & String  & -- \\
Experimental          & Boolean & -- \\
DOI                   & String  & -- \\
\end{tabular}
\end{ruledtabular}
\end{table}

\subsection{\label{sec:level2}Fine-tuning LLMs for Superconductor Classification}

Rapid and accurate identification of superconductors remains a fundamental challenge in materials science. Traditional machine learning methods rely on manually designed features, whereas large language models (LLMs) represent a major shift by learning useful representations directly from raw text or chemical compositions through attention mechanisms, removing the need for explicit feature engineering. This approach enables LLMs to infer complex relationships from natural language descriptions or chemical formulas, making them an appealing alternative to conventional classifiers.

In recent years, numerous open-source LLMs have become readily accessible, providing an unprecedented opportunity to benchmark their capabilities for scientific discovery tasks. However, as our initial tests show, base versions of popular LLMs---such as Mistral-7B, Llama 3.1-8B, Qwen3-14B, Qwen3-2507-4B, and Phi4-14B---are biased toward predicting all compounds as superconductors when queried directly with raw chemical compositions. This bias likely stems from imbalances in the scientific literature: positive (superconductor) results are reported far more frequently than negative (non-superconductor) outcomes, resulting in pre-trained models with limited exposure to the full range of material classes.

To construct a more balanced and representative dataset for supervised fine-tuning, we merged our LLM-generated superconductor database  with the SuperCon resource and further supplemented the negative class with experimentally verified non-superconducting magnetic compounds from the NEMAD database, consistent with established practices~\cite{konno2021deep}. This integration resulted in a comprehensive dataset of 45,116 entries, providing a robust foundation for training and evaluation.

Each LLM was then fine-tuned for binary classification (outputting 1 for superconductors and 0 for non-superconductors) using an 80:20 train-test split, with supervised learning implemented via the Unsloth package. Importantly, we explicitly instructed the models to output only the class label (as a number) in response to a given chemical composition. This minimizes the generation of extraneous tokens and ensures that any deviation from the target class is directly penalized in the cross-entropy loss, facilitating clearer optimization and evaluation during training.

For benchmarking, we trained XGBoost and random forest classifiers using 84-dimensional elemental proportion vectors and other composition-based features obtained from the \texttt{pymatgen} library~\cite{ong2013python,ward2016general}.

Table~\ref{table:classification} provides a comprehensive comparison of statistical performance across all models, including precision, recall, F1-score, and accuracy. All fine-tuned LLMs achieve high accuracy, clustering around 91\%, with Qwen3-14B delivering the best overall F1-score (0.919) and accuracy (0.911). These results are comparable to, and in some metrics surpass, the performance of established feature-based models.

\begin{figure*}[ht]
\includegraphics[width=\textwidth]{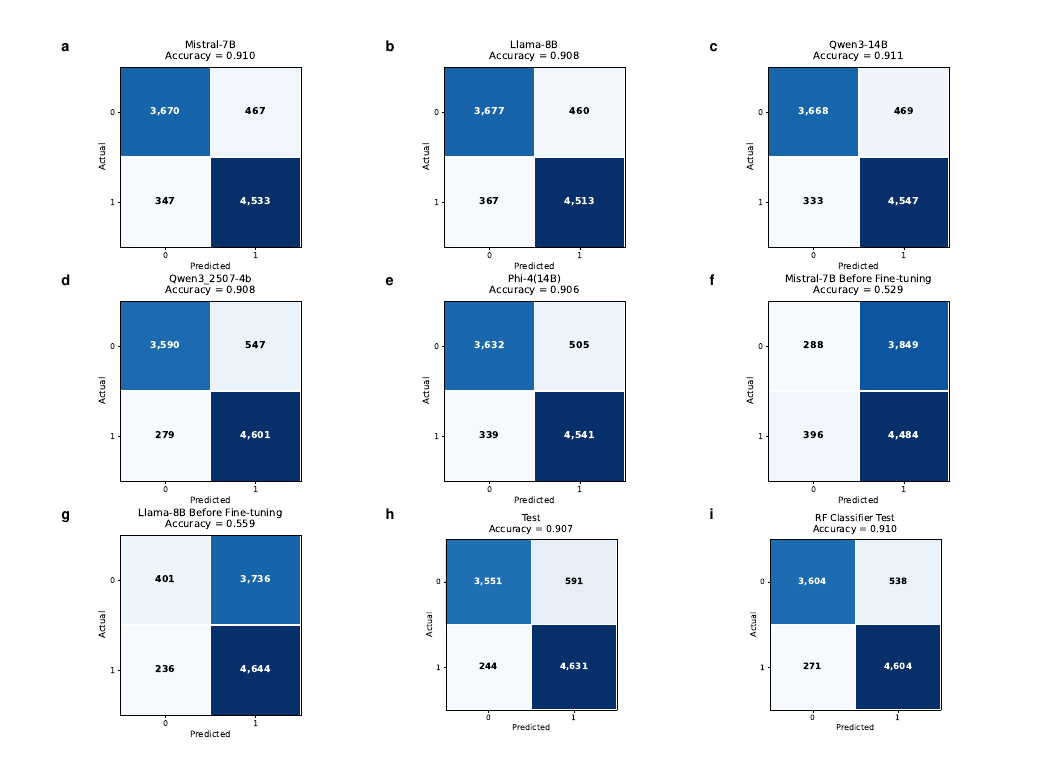}
\caption{\label{fig:confusion_matrix}
\textbf{Confusion matrices for SC/NSC classification using fine-tuned and base LLMs, compared with feature-based classifiers.}
\textbf{(a–e)} show the performance of fine-tuned Mistral-7B, Llama-8B, Qwen3-14B, Qwen3-2507-4B, and Phi-14B, all achieving accuracies of approximately 0.90--0.91.
\textbf{(f–g)} display the corresponding base (pre--fine-tuning) models, which perform substantially worse (accuracies 0.53--0.56), demonstrating the critical role of supervised instruction tuning.
\textbf{(h)} shows the held-out test performance of an XGBoost classifier (accuracy 0.907).
\textbf{(i)} provides the Random Forest baseline (accuracy 0.910).
Together, these results indicate that fine-tuned LLMs reach classification performance comparable to established feature-based models while significantly improving over their pre-trained counterparts.}
\end{figure*}

Figure~\ref{fig:confusion_matrix} visualizes the confusion matrices for each approach. Fine-tuned LLMs (Fig.~\ref{fig:confusion_matrix}a–e) exhibit strong predictive performance, with the majority of predictions aligning along the diagonal and minimal class bias. In contrast, the base models (Fig.~\ref{fig:confusion_matrix}f, g) display pronounced bias towards the superconductor class, underlining the necessity of targeted fine-tuning. The feature-based models (Fig.~\ref{fig:confusion_matrix}h, i) also perform well, but do not substantially outperform the fine-tuned LLMs.

Despite these encouraging results, we note that some compounds are consistently misclassified across different models---particularly those with ambiguous or borderline properties. Analyzing these challenging cases may provide further insights into both the limitations of current classification approaches and the underlying complexity of superconductivity as a materials property.

Overall, these findings demonstrate that with appropriate supervised fine-tuning, open-source LLMs can match or exceed the predictive power of traditional feature-based classifiers, while leveraging only raw compositional input. This capability holds significant promise for accelerating the discovery and screening of novel superconducting materials.

\begin{table}[ht]
\caption{\label{table:classification}
\textbf{Performance comparison of different classification models.}}
\begin{ruledtabular}
\begin{tabular}{lcccc}
Model & Precision & Recall & F1-score & Accuracy \\
\hline
Mistral-7B         & 0.907 & 0.929 & 0.918 & 0.910 \\
Llama 3.1-8B       & 0.908 & 0.925 & 0.916 & 0.909 \\
Qwen3-14B          & 0.907 & 0.932 & \textbf{0.919} & \textbf{0.911} \\
Qwen3-2507-4B      & 0.894 & 0.943 & 0.918 & 0.908 \\
Phi4-14B           & 0.900 & 0.931 & 0.915 & 0.906 \\
Mistral-7B Base    & 0.538 & 0.919 & 0.679 & 0.529 \\
Llama-8B Base      & 0.554 & \textbf{0.952} & 0.701 & 0.559 \\
XGB Classifier     & 0.911 & 0.904 & 0.906 & 0.907 \\
RF Classifier      & \textbf{0.913} & 0.907 & 0.909 & 0.910 \\
\end{tabular}
\end{ruledtabular}
\end{table}

\subsection{\label{sec:level2}Fine-tuning LLMs for Predicting the Critical Temperature of Superconductors}

\begin{figure*}[ht]
\includegraphics[width=\textwidth]{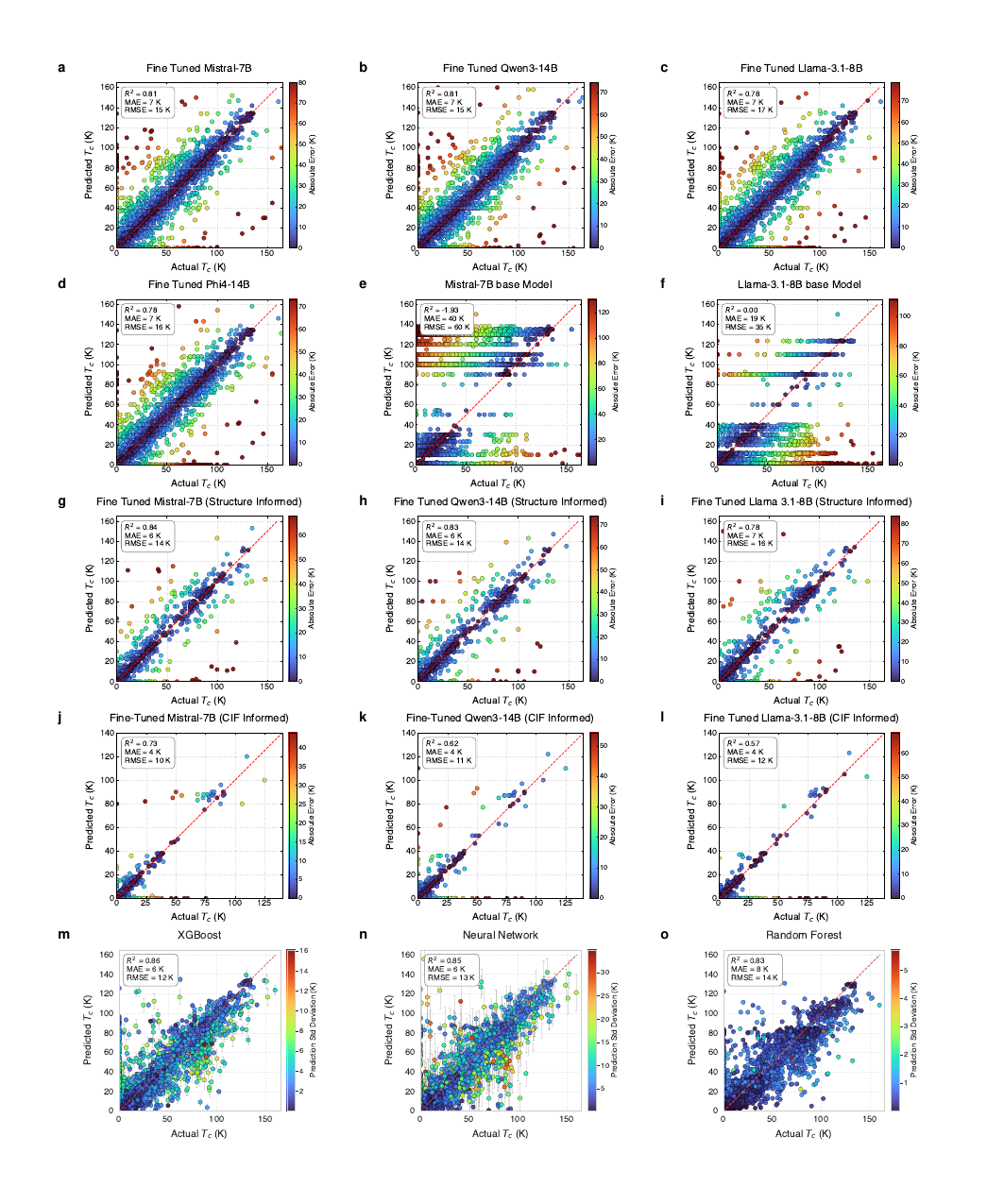}
\caption{\label{fig:tc_prediction}
\textbf{Performance of fine-tuned LLMs, base models, and feature-based baselines for superconducting transition-temperature ($T_c$) regression.}
\textbf{(a–d)} Fine-tuned Mistral-7B, Qwen3-14B, Llama-3.1-8B, and Phi4-14B using composition-only inputs.
\textbf{(e–f)} Corresponding base (pre--fine-tuning) models evaluated on the same test set, showing substantially lower predictive accuracy.
\textbf{(g–i)} Models fine-tuned with composition plus crystal system and space group, reflecting improved structure-aware prediction.
\textbf{(j–l)} Models trained on full CIF-informed inputs, where longer serialized structures and reduced dataset size lead to degraded performance.
\textbf{(m–o)} Feature-based regressors—XGBoost, a neural-network ensemble, and Random Forest—trained on engineered compositional descriptors.}
\end{figure*}

Based on the classification model, we can proceed to the prediction of critical temperatures for those classified into superconducting category. 
Traditional approaches to \(T_c\) prediction---both classical machine learning and deep learning---typically rely on manually engineered features derived from composition or crystal structure. In contrast, large language models (LLMs) can potentially infer complex chemical–physical relationships directly from text or symbolic representations through attention mechanisms, removing the need for handcrafted descriptors.

Here, we fine-tuned a set of open-source transformer-based LLMs---Mistral-7B, Llama 3.1-8B, Qwen3-14B, and Phi4-14B---for \(T_c\) regression tasks. Each model was trained in a supervised manner using chemical composition as input and the experimental \(T_c\) value as output. To ensure consistency, the models were fine-tuned using the same 80:20 train–test split described earlier. The base models were also evaluated on the same test set for comparison. Figure~\ref{fig:tc_prediction}a–d shows the results for the fine-tuned models, whereas Figure~\ref{fig:tc_prediction}e–f illustrates the performance of the base versions. The base LLMs exhibited poor regression performance, with \(R^2\) values close to zero or even negative, and high mean absolute errors (MAE~\(\approx\)~19–40~K), indicating that pre-trained models fail to generalize the quantitative mapping between composition and \(T_c\). Their predictions cluster heavily around low-\(T_c\) regions (typically 10–30~K), suggesting that these models may have primarily encountered BCS-type superconductors during pre-training.

\begin{table}[ht]
\caption{\label{table:tc_prediction}
\textbf{Performance comparison of different models for $T_c$ prediction.}}
\begin{ruledtabular}
\begin{tabular}{lccccc}
Model & Dataset & Size & $R^2$ & MAE & RMSE \\
\hline
Mistral-7B         & C         & 26k & 0.81 & 7K  & 15K \\
Mistral-7B         & C+CS+SG   & 13k & 0.84 & 6K  & 14K \\
Mistral-7B         & C+Cif     & 7k  & 0.73 & 4K  & 10K \\

Qwen3-14B          & C         & 26k & 0.81 & 7K  & 15K \\
Qwen3-14B          & C+CS+SG   & 13k & 0.83 & 6K  & 14K \\
Qwen3-14B          & C+Cif     & 7k  & 0.62 & 4K  & 11K \\

Llama 3.1-8B       & C         & 26k & 0.78 & 7K  & 17K \\
Llama 3.1-8B       & C+CS+SG   & 13k & 0.78 & 7K  & 16K \\
Llama 3.1-8B       & C+Cif     & 7k  & 0.57 & 4K  & 12K \\

Phi4-14B           & C         & 26k & 0.78 & 7K  & 16K \\

Mistral-7B Base    & --        & --  & -1.93 & 40K & 60K \\
Llama-8B Base      & --        & --  & 0.00  & 19K & 35K \\

XGBoost            & C         & 26k & 0.86 & 6K  & 12K \\
Neural Network     & C         & 26k & 0.85 & 6K  & 13K \\
Random Forest      & C         & 26k & 0.83 & 8K  & 14K \\
\end{tabular}
\end{ruledtabular}
\end{table}

 After fine-tuning, all models showed substantial improvement: Mistral-7B and Qwen3-14B achieved the highest \(R^2\) values of 0.81, while Llama~3.1-8B and Phi4-14B reached 0.78 (Table~\ref{table:tc_prediction}). The corresponding MAE values ranged between 6–7~K, confirming that the fine-tuned models captured both low- and intermediate-\(T_c\) regimes with reasonable accuracy. A comprehensive summary of the performance metrics for all models is provided in Table~\ref{table:tc_prediction}.

To examine the effect of incorporating structural information, we trained additional models where crystal system (CS), space group (SG), and applied pressure data were appended to the chemical composition (Figure~\ref{fig:tc_prediction}g–i). The inclusion of these features led to a modest yet consistent improvement in model performance: for Mistral-7B, \(R^2\) increased from 0.81 to 0.84, and for Qwen3-14B, from 0.81 to 0.83. However, the Llama~3.1-8B model showed no significant gain, maintaining \(R^2 = 0.78\), suggesting that the benefit of structural metadata depends on the model architecture and how efficiently it integrates auxiliary context.

We further extended the dataset by incorporating full crystallographic information files (CIFs) into the model input. Although our primary database contains only space-group and crystal-system labels, we obtained complete CIFs by matching our compositions and space groups against two external crystallographic resources: the Crystallography Open Database (COD)~ \cite{Grazulis2012,Grazulis2009} and the Materials Project~\cite{jain2013commentary}. To further increase coverage, we integrated entries from the open-access 3DSC database~\cite{sommer20233dsc}, which provides CIFs for known superconductors. Combining these sources yielded a curated subset of approximately 7,000 materials with full structural information.

Including CIFs significantly increases sequence length and token complexity, since each file encodes lattice parameters, atomic coordinates, Wyckoff positions, and site occupations. The CIF-informed fine-tuned models (Figure~\ref{fig:tc_prediction}j–l) achieved $R^2$ values of 0.73 (Mistral-7B), 0.62 (Qwen3-14B), and 0.57 (LLaMA~3.1-8B), with mean absolute errors remaining low ($\approx 4$~K). However, their overall performance decreased compared to the composition-only and CS+SG-informed models. This reduction can be attributed to two factors: (i) CIF inputs produce substantially longer and more complex token sequences, making optimization more challenging, especially under parameter-efficient fine-tuning; and (ii) the CIF-matched subset is considerably smaller and biased toward low-$T_c$ compounds, resulting in reduced data diversity and fewer high-temperature examples for the model to learn from.

For benchmarking, we trained traditional feature-based regression models---XGBoost, neural networks, and random forests---using composition-derived features generated via the \texttt{pymatgen} library~\cite{ong2013python,ward2016general}. As shown in Figure~\ref{fig:tc_prediction}m–o and Table~\ref{table:tc_prediction}, these models achieved \(R^2\) scores of 0.83–0.86 and MAE of 6–8~K, comparable to or slightly better than the fine-tuned LLMs. Despite this, the fine-tuned LLMs are remarkable in that they learned quantitative structure–property relationships directly from raw chemical formulas, without any feature engineering.

In summary, fine-tuning large language models substantially enhances their capability to predict the critical temperature of superconductors. While current LLM-based approaches achieve accuracy close to feature-based machine learning, their ability to ingest unstructured textual and structural data positions them as powerful, general-purpose tools for data-driven materials discovery. Future optimization---through multimodal fusion of text, composition, and structure representations---may further bridge the gap between language-based and physics-informed prediction models.

\clearpage

\subsection{\label{sec:level2}Inverse Design of Superconductors Using Fine-tuned LLMs}

Beyond forward prediction tasks, a key advantage of large language models (LLMs) is their ability to perform inverse design---generating candidate materials that satisfy desired physical properties. In the context of superconductivity, this task involves proposing chemically and structurally plausible superconducting compositions conditioned on a target critical temperature (\(T_c\)) and, optionally, a specified crystal system. Achieving reliable inverse design is challenging, particularly due to the scarcity of complete structural information (e.g., CIF files) across known superconductors. As our forward \(T_c\) models already demonstrated limited improvement when full crystallographic inputs were used, generating complete CIFs is currently infeasible with the available data. Therefore, we restricted the inverse design task to predicting chemical compositions and, when possible, their corresponding space groups.

To construct a suitable training set, we generated two types of instruction–response pairs. In the first dataset, the model was given a target \(T_c\) and a crystal system, and was instructed to output a plausible superconducting composition together with its space group and, when applicable, the applied pressure. In the second dataset, the model was conditioned only on a target \(T_c\) and asked to generate a corresponding superconducting composition. Combining both formats resulted in approximately 37,000 instruction–response examples in Alpaca style. Based on its strong performance in both classification and regression, Qwen3-14B was selected as the model for fine-tuning on the inverse design task.

To evaluate the generative performance, we constructed a grid of target conditions consisting of temperatures from 5 to 200~K in 1~K increments and all seven crystal systems, yielding 1,568 unique prompts. For each prompt, we sampled the model ten times to assess generative diversity, resulting in a total of 15,680 model queries. The same instruction formats used during training were applied during inference.

\begin{figure}[ht]
\includegraphics[width=0.50\textwidth]{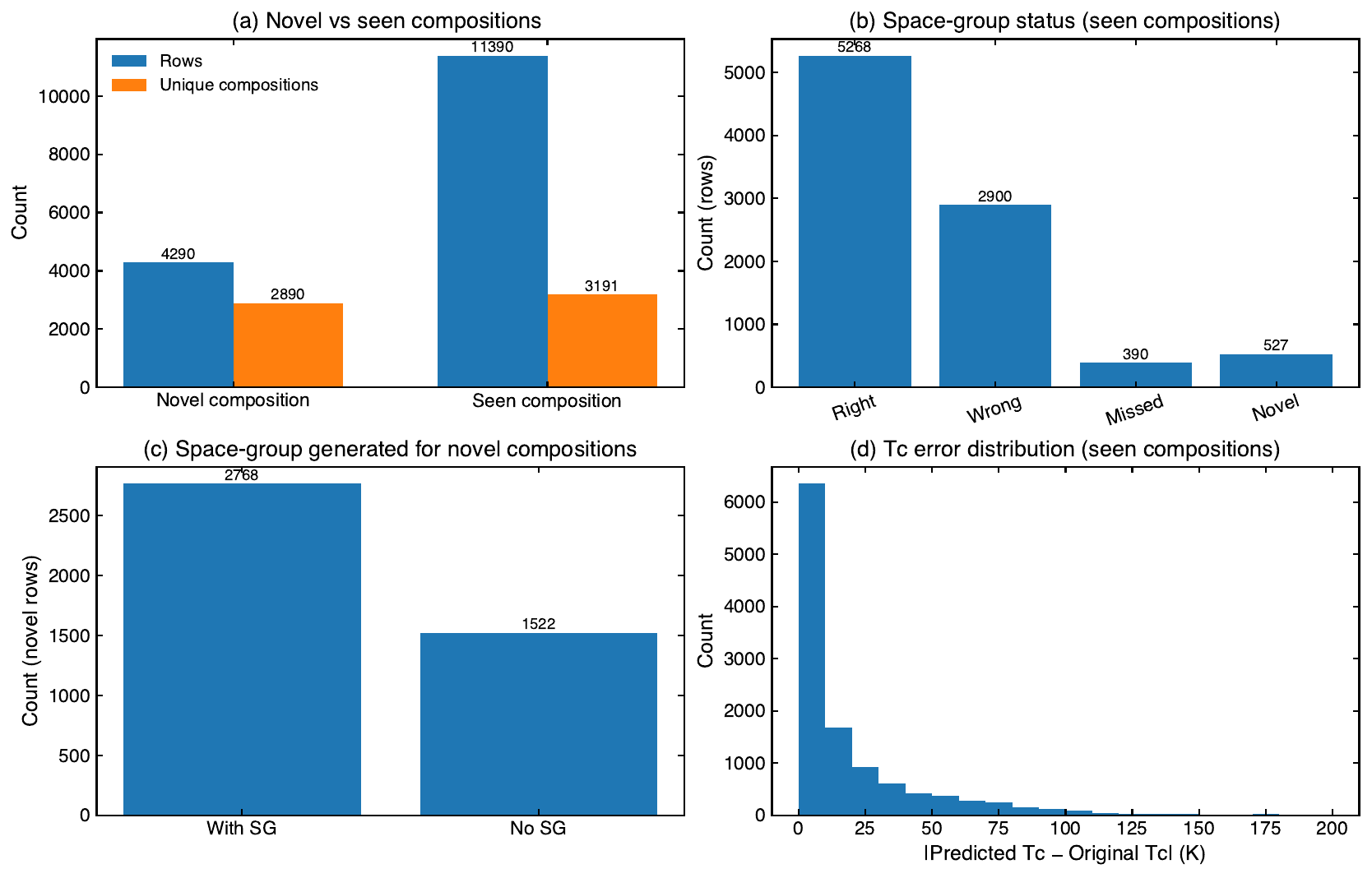}
\caption{\label{fig:inverse_design}
\textbf{Evaluation of the fine-tuned Qwen3-14B model for inverse design of superconductors.}
\textbf{(a)} Number of generated compositions that are novel (not present in the training set) versus those that match compositions present in the training set; among 15,680 model samples, 4,290 generated formulas were novel, of which 2,890 were unique compositions.
\textbf{(b)} Space-group accuracy for generated compositions that match entries in the training data. The model reproduced the correct space group in 5,268 cases, assigned an alternative space group in 2,900 cases, omitted the space group in 390 cases, and proposed a previously unseen space group in 527 cases.
\textbf{(c)} Space-group generation for novel compositions. Of the 4,290 novel rows, 2,768 included both a chemical formula and a predicted space group, while 1,522 contained only a formula.
\textbf{(d)} Distribution of $\lvert T_c^{\mathrm{target}} - T_c^{\mathrm{original}} \rvert$ for generated compositions that match known materials. Most reproduced compositions lie within 20~K of the target value, indicating that the model preferentially selects materials with experimentally reported $T_c$ values close to the specified target.}
\end{figure}

Figure~\ref{fig:inverse_design} summarizes the performance of the fine-tuned inverse design model. Across all queries, the model produced 4,290 compositions not present in the training set, of which 2,890 were unique (Fig.~\ref{fig:inverse_design}a). This indicates that the model possesses meaningful generative capability and is not merely memorizing its training distribution. Among the novel compositions, 2,768 were generated with both a chemical formula and an associated space group (Fig.~\ref{fig:inverse_design}c), demonstrating that the model can design new materials while simultaneously proposing plausible structural information.

For generated compositions that matched entries in the training set, we evaluated the model’s ability to reproduce the correct space-group information. As shown in Fig.~\ref{fig:inverse_design}b, the model generated the correct space group in 5,268 cases. In other 2,900 cases, however, it assigned a different space group despite the composition being identical to one in the training set. Although we label these as ``wrong'', such differences do not necessarily imply an error, as multiple space groups can be crystallographically feasible for the same composition. In addition, the model omitted the space group entirely for 390 entries even though the training data provided one, while in 527 cases it proposed a novel space group not present in the training set. The latter outcome is noteworthy, as it may reflect the model’s ability to explore alternative symmetry assignments rather than simple memorization. Together, these behaviours illustrate both the capabilities and the current limitations of the model’s structural reasoning.

We further evaluated how well the generated compositions satisfy the requested \(T_c\) values for cases where the model reproduced a composition that was already present in the training set. For each such entry, we retrieved its experimental \(T_c\) from the database and computed the absolute deviation \(|T_c^{\mathrm{target}} - T_c^{\mathrm{original}}|\) between the database value and the target value used in the prompt. Figure~\ref{fig:inverse_design}d shows the distribution of this error. The majority of cases lie within 20~K of the requested \(T_c\), even though the target temperatures were sampled uniformly from 5--200~K. This indicates that, when it selects compositions from the known data manifold, the model often chooses materials whose known \(T_c\) is close to the specified target.

Although the present inverse design model does not attempt to generate full crystallographic information, it nevertheless serves as a valuable tool for producing synthetic superconducting compositions and enriching existing datasets. As machine learning models for superconductivity continue to advance, such domain-specific synthetic data may help alleviate current data bottlenecks, improve model generalization, and ultimately support the development of more reliable and comprehensive inverse design frameworks.

\subsection{Identification of Novel Superconductor Candidates from External Databases}

Several large inorganic materials databases---derived from density functional theory calculations or deep-learning–based generative models---contain extensive information on crystal structures and thermodynamic stability but lack any information regarding superconductivity. To explore whether our models can identify previously unrecognized superconducting candidates within these repositories, we screened the GNoME database \cite{merchant2023scaling}.


To avoid rediscovering known materials, all compositions present in our curated superconductor dataset were first removed from these external datasets. The remaining entries were reformatted to match the input requirements of both our fine-tuned LLM models and the feature-based machine learning models. Screening proceeded in two stages. First, all classification models were applied to label each composition as either superconducting or non-superconducting. Only entries consistently classified as superconductors were retained for further analysis. In the second stage, we applied our regression models---including both fine-tuned LLM regressors and feature-based regressors---to estimate the corresponding critical temperatures.

Through this multi-step screening workflow, we identified a set of previously unreported candidate superconductors with predicted critical temperatures exceeding 10~K for all models. Table~\ref{table:supp_tc} lists the composition appeared as superconducting by the classification ensemble together with their predicted \(T_c\) values. While these predictions have not yet been experimentally verified, they offer plausible high-\(T_c\) candidates emerging from large external materials databases. These compositions may serve as starting points for future experimental exploration or more detailed computational investigations.

\section{\label{sec:level1}Conclusions}

In this work, we demonstrated an end-to-end framework that integrates large language models into the full pipeline of data extraction, curation, and property prediction for superconducting materials. Using a prompt-engineered extraction workflow, we assembled a comprehensive experimental database of 78,203 records covering 19,058 unique compositions, each annotated with chemical, structural, superconducting transition temperature, and magnetic-field information. The assembled dataset enlarges the available experimental record on superconductors and incorporates property information that is not contained in existing repositories.

We fine-tuned several open-source language models to perform superconductivity classification, critical-temperature regression, and inverse design. The resulting classification models achieve performance comparable to---and in some cases exceeding---traditional feature-based approaches, while also significantly improving upon their base versions. Although feature-engineered models remain the top performers for $T_c$ regression, the fine-tuned LLMs closely approach their accuracy while substantially outperforming their base versions and learning chemically and structurally meaningful trends directly from text inputs.

Beyond prediction, we showed that fine-tuned models can generate chemically plausible compositions for inverse-design tasks conditioned on a target $T_c$. 
Separately, by applying the forward-prediction models to external materials databases, we identified compounds with predicted transition temperatures above 10~K, demonstrating the utility of LLM-based screening for uncovering new superconducting candidates

Looking ahead, the framework introduced here can be extended in several ways.
Larger-scale language models may further improve regression accuracy, and the incorporation of physics-informed constraints offers a potential route toward more reliable predictions. As broader crystallographic data become available, integrating full structural representations may also enhance both forward prediction and inverse-design performance. These directions highlight opportunities for building increasingly robust, physically grounded LLM-based tools for superconductivity research. A portal on the NEMAD database has been created for the whole community to contribute their recent superconductor data. The database could be a scalable foundation for future data-driven discovery in quantum materials.


\section*{Acknowledgments}
This work was supported by the Office of Basic Energy Sciences, Division of Materials Sciences and Engineering, U.S. Department of Energy, under Award No. DE-SC0020221. 

\section*{Data availability}

All resources developed in this work are publicly available. The superconductivity
database used for model training and evaluation is accessible at
\href{https:www.nemad.org}{www.nemad.org}. All fine-tuned large language models (LLMs), along with
downloadable links and instructions for their use, are provided in a GitHub
repository at \href{https://github.com/sumanitani/LLM_for_Superconductors}{github.com/sumanitani/LLM for Superconductors}.


\onecolumngrid   
\section*{Appendix}

\begin{longtable}{l l l c c c c c c}
\caption{\textbf{Full predicted $T_c$ (K) values from different models for screened high-$T_c$ materials.}}\\
\label{table:supp_tc} \\
\toprule
MaterialId & Composition & Space Group & Mistral(C+S) & Mistral(C) & Qwen3(C+S) & Qwen3(C) & XGB & NN \\
\midrule
\endfirsthead

\caption[]{(Continued)}\\
\toprule
MaterialId & Composition & Space Group & Mistral(C+S) & Mistral(C) & Qwen3(C+S) & Qwen3(C) & XGB & NN \\
\midrule
\endhead

\midrule
\multicolumn{9}{r}{\textit{Continued on next page}}\\
\endfoot

\bottomrule
\endlastfoot

01ed6ada43 & Li4NbN3 & Ibca & 15.0 & 18.0 & 10.0 & 15.3 & 20.35 $\pm$ 3.9 & 16.19 $\pm$ 10.72 \\
05e0b6f484 & Na2LiZr(H6Ir)2 & P-42m & 14.0 & 14.0 & 10.0 & 14.0 & 60.09 $\pm$ 8.77 & 24.28 $\pm$ 16.5 \\
0fe0fc19f1 & YZr7H30 & R3 & 115.0 & 144.0 & 145.0 & 144.0 & 42.53 $\pm$ 7.36 & 48.49 $\pm$ 28.46 \\
10d228680c & Ca9AcTaNbN10 & P1 & 10.0 & 13.0 & 13.0 & 14.0 & 13.22 $\pm$ 2.08 & 19.44 $\pm$ 15.25 \\
1245da3d42 & Sc5TiH18 & C2 & 135.0 & 117.0 & 144.0 & 101.0 & 39.48 $\pm$ 6.38 & 37.87 $\pm$ 15.65 \\
137be06481 & Rb5Ba(MoH9)3 & Amm2 & 117.0 & 100.0 & 100.0 & 100.0 & 48.46 $\pm$ 8.44 & 58.08 $\pm$ 37.3 \\
20ec23fb02 & SrLi(H3Pt)2 & Fm-3m & 11.0 & 11.0 & 20.0 & 14.0 & 27.51 $\pm$ 2.9 & 13.77 $\pm$ 11.26 \\
2abe6657b1 & SrLi4 & P6\_3/mmc & 11.0 & 34.0 & 11.05 & 13.5 & 11.76 $\pm$ 5.89 & 20.98 $\pm$ 22.11 \\
2ed395579d & YSc5H18 & Cm & 145.0 & 118.0 & 110.0 & 146.0 & 39.47 $\pm$ 5.89 & 39.67 $\pm$ 13.3 \\
2f265922e7 & Na4Cu2(TcH9)3 & Amm2 & 13.0 & 11.0 & 10.05 & 50.0 & 72.66 $\pm$ 15.6 & 49.5 $\pm$ 26.13 \\
330c818ac4 & ThNp6PaH30 & R3 & 11.0 & 14.3 & 34.05 & 14.0 & 31.47 $\pm$ 13.71 & 12.82 $\pm$ 12.55 \\
3519e6cf7a & CaTmMg2(B6Os)2 & P2/m & 32.0 & 35.0 & 33.0 & 38.0 & 25.2 $\pm$ 3.71 & 30.02 $\pm$ 15.86 \\
390694f010 & Li12Ca6Hg4N & Im-3m & 30.0 & 40.0 & 12.0 & 12.0 & 30.1 $\pm$ 7.07 & 52.35 $\pm$ 27.99 \\
399df12841 & NaB12PC & Ima2 & 10.0 & 12.0 & 10.0 & 19.0 & 21.36 $\pm$ 5.09 & 21.83 $\pm$ 11.4 \\
3cbabdd56f & NaLi4Ca(RhN)4 & P4/m & 24.0 & 39.0 & 25.0 & 12.0 & 28.35 $\pm$ 6.06 & 23.96 $\pm$ 16.23 \\
3f9493cc37 & KSrH9W & P-62m & 140.0 & 152.0 & 100.0 & 146.0 & 50.13 $\pm$ 11.55 & 20.67 $\pm$ 30.73 \\
438be15eb0 & Ba5Ca(VH9)3 & Amm2 & 75.0 & 105.0 & 100.0 & 140.0 & 51.89 $\pm$ 9.23 & 66.93 $\pm$ 45.57 \\
4be360c065 & ZrSc5H18 & C2 & 110.0 & 114.0 & 148.0 & 145.0 & 31.15 $\pm$ 6.11 & 29.52 $\pm$ 11.22 \\
5050b5f8a4 & Ba(Sr2Cu3)2 & Amm2 & 30.0 & 45.0 & 60.0 & 50.0 & 55.76 $\pm$ 7.69 & 45.32 $\pm$ 31.25 \\
544bc02c36 & HfMg7(B6Ru)4 & Pm & 11.0 & 24.0 & 25.05 & 20.0 & 23.22 $\pm$ 1.36 & 28.5 $\pm$ 6.86 \\
56d64c3947 & Sc5TaH18 & C2 & 100.0 & 115.0 & 108.0 & 100.0 & 26.97 $\pm$ 2.63 & 18.45 $\pm$ 23.34 \\
57c01b9a23 & CaB9N & R-3m & 12.0 & 10.5 & 10.05 & 14.5 & 20.95 $\pm$ 4.29 & 20.47 $\pm$ 12.45 \\

5bdcd48919 & Sc5PaH18 & C2 & 110.0 & 115.0 & 103.0 & 143.0 & 26.97 $\pm$ 3.21 & 30.33 $\pm$ 22.45 \\
660ca97914 & NaZr(ReH9)2 & C2/m & 45.0 & 14.0 & 11.05 & 14.0 & 73.88 $\pm$ 7.39 & 106.26 $\pm$ 54.09 \\
6a0c149e22 & Mg5Cd(TcH7)2 & P-6m2 & 10.0 & 16.6 & 30.05 & 34.0 & 56.31 $\pm$ 5.88 & 34.0 $\pm$ 19.34 \\
6b825c87e0 & KLa2(H6Ir)2 & P-3m1 & 150.0 & 138.0 & 138.0 & 140.0 & 37.55 $\pm$ 3.44 & 16.53 $\pm$ 15.92 \\
6ff2843171 & Mg5Cu(TcH7)2 & P-6m2 & 10.0 & 15.8 & 30.05 & 30.0 & 54.7 $\pm$ 5.57 & 33.61 $\pm$ 22.27 \\
710a366bd7 & Li3CaAlMo3H20 & P1 & 11.0 & 30.0 & 40.0 & 40.0 & 60.63 $\pm$ 11.72 & 83.41 $\pm$ 17.63 \\
789a40e4d3 & Sr6Li12Tl4N & Im-3m & 40.0 & 11.0 & 10.05 & 12.0 & 37.78 $\pm$ 7.75 & 43.18 $\pm$ 24.12 \\
7da7d6131f & TmSc5H18 & C2 & 110.0 & 108.0 & 103.0 & 145.0 & 37.92 $\pm$ 5.95 & 32.96 $\pm$ 27.27 \\
805a8fc9a8 & KRb(BC7)2 & C2 & 25.0 & 27.0 & 10.05 & 15.0 & 24.17 $\pm$ 2.62 & 21.39 $\pm$ 16.28 \\
84133d870f & Ba2(SrCu2)3 & Cmmm & 80.0 & 50.0 & 80.0 & 70.0 & 58.59 $\pm$ 10.84 & 57.52 $\pm$ 39.22 \\
896c3355a3 & Sr4La2(MoH9)3 & Amm2 & 100.0 & 65.0 & 100.0 & 100.0 & 40.15 $\pm$ 10.16 & 37.73 $\pm$ 29.14 \\
8b3102b39f & RbBaH9W & P-62m & 110.0 & 150.0 & 100.0 & 146.0 & 46.22 $\pm$ 12.13 & 17.34 $\pm$ 22.79 \\
8bc068ef0c & SrLaCuN2 & I4mm & 20.0 & 40.0 & 30.0 & 20.0 & 17.77 $\pm$ 3.54 & 31.81 $\pm$ 18.27 \\
8d727bf7e3 & CsAc(BH4)3 & P2\_12\_12 & 100.0 & 117.0 & 100.0 & 100.0 & 67.63 $\pm$ 4.72 & 68.91 $\pm$ 20.88 \\
9a1ea7b085 & NaCa3NbN4 & P1 & 12.0 & 11.5 & 15.0 & 14.0 & 19.22 $\pm$ 3.65 & 11.21 $\pm$ 6.97 \\
9cfe4ee521 & SrLi3CaNbN4 & Pnnm & 12.0 & 32.0 & 15.0 & 16.0 & 27.35 $\pm$ 8.11 & 20.78 $\pm$ 15.54 \\
9da8e6e9e0 & Mg2TlH8Rh & P4\_2/mmc & 30.0 & 80.0 & 54.0 & 100.0 & 69.36 $\pm$ 3.5 & 71.63 $\pm$ 28.06 \\
9dffc6eb18 & PrMg(ReH9)2 & C2/m & 70.0 & 20.0 & 35.05 & 14.0 & 56.18 $\pm$ 6.97 & 50.68 $\pm$ 35.05 \\
a0005d43be & H8Se3S & P2\_1/m & 130.0 & 110.0 & 115.0 & 114.0 & 89.04 $\pm$ 10.55 & 86.63 $\pm$ 22.05 \\
a0da37ef5c & Mg8B23Ru4C & Pm & 37.0 & 37.6 & 34.0 & 29.0 & 26.14 $\pm$ 0.8 & 32.58 $\pm$ 6.51 \\
a1f84501d1 & YB9N & R-3m & 12.0 & 10.5 & 10.0 & 14.0 & 18.82 $\pm$ 6.69 & 11.65 $\pm$ 6.74 \\
a356cabe64 & Ba6CuC2N7 & P2\_12\_12 & 40.0 & 50.0 & 30.0 & 55.0 & 29.16 $\pm$ 5.86 & 18.31 $\pm$ 20.56 \\
a454484f40 & AcAlH6 & R-3m & 115.0 & 100.0 & 100.0 & 100.0 & 44.51 $\pm$ 5.45 & 58.35 $\pm$ 16.8 \\
a87ede8637 & YMgB6Os & Pbam & 11.0 & 12.0 & 35.0 & 29.0 & 27.2 $\pm$ 4.18 & 17.2 $\pm$ 8.53 \\
aa5038f61a & Rb3Ba2BiO6 & Fddd & 25.0 & 13.0 & 11.0 & 32.0 & 51.42 $\pm$ 5.46 & 50.36 $\pm$ 30.1 \\
ad5ddf6bd9 & Th7NpH24 & Pm-3 & 13.9 & 10.1 & 30.05 & 14.4 & 17.88 $\pm$ 10.27 & 11.99 $\pm$ 9.65 \\
b1576dddcc & BPH6 & P2\_1 & 115.0 & 100.0 & 100.0 & 104.0 & 97.98 $\pm$ 15.95 & 111.11 $\pm$ 17.66 \\
b2d9091623 & LiCa2C3F & P4/mbm & 12.0 & 11.5 & 10.05 & 11.5 & 21.33 $\pm$ 5.4 & 12.67 $\pm$ 9.99 \\
b3759b4259 & LaMg5(TcH7)2 & P-6m2 & 30.0 & 14.0 & 34.05 & 47.0 & 45.51 $\pm$ 7.81 & 27.59 $\pm$ 21.32 \\
b850609bf4 & CeTh6PaH30 & P1 & 11.2 & 16.6 & 11.05 & 14.5 & 19.48 $\pm$ 11.46 & 26.62 $\pm$ 22.52 \\
bb845e364a & Sr4Ca4Ta(NbN4)3 & P1 & 30.0 & 35.0 & 10.05 & 13.0 & 13.55 $\pm$ 4.98 & 15.91 $\pm$ 13.01 \\
bc769d74fc & Li6TcN4 & P4\_2/nmc & 12.0 & 14.4 & 25.05 & 14.5 & 26.11 $\pm$ 6.57 & 20.87 $\pm$ 13.19 \\
be00b8114b & NaLi(B13C2)2 & Imm2 & 50.0 & 38.0 & 20.0 & 30.0 & 21.27 $\pm$ 5.33 & 17.9 $\pm$ 9.62 \\
c3ac0d8b9a & Sr4Mg2AlTcH20 & P1 & 37.0 & 18.0 & 24.0 & 20.0 & 50.0 $\pm$ 11.19 & 66.33 $\pm$ 30.95 \\
c4b61bdcce & Mg8B23Os4C & Pm & 37.0 & 38.0 & 35.0 & 29.0 & 26.26 $\pm$ 0.69 & 25.27 $\pm$ 5.93 \\
cfea2d0e69 & Sr3Ca3(TcH7)2 & P3m1 & 100.0 & 110.0 & 60.0 & 100.0 & 63.83 $\pm$ 8.06 & 32.76 $\pm$ 24.87 \\
d45484b3bc & Th3ScH15 & Cc & 110.0 & 109.0 & 105.0 & 100.0 & 31.21 $\pm$ 7.5 & 30.53 $\pm$ 21.49 \\
d643242a4b & LiNbH4 & P4/mmm & 15.0 & 42.0 & 33.0 & 14.0 & 43.5 $\pm$ 5.87 & 34.25 $\pm$ 20.59 \\
de0c587e9e & Ba2Ca7(BN2)6 & R-3m & 44.0 & 75.0 & 50.0 & 55.0 & 15.18 $\pm$ 4.64 & 14.05 $\pm$ 12.69 \\
dffcd0eade & B(HS)3 & P2\_1/c & 100.0 & 118.0 & 100.0 & 98.0 & 64.81 $\pm$ 9.56 & 55.44 $\pm$ 18.57 \\
e259827901 & SrAc3H11 & I4mm & 115.0 & 100.0 & 103.0 & 100.0 & 34.01 $\pm$ 9.04 & 14.66 $\pm$ 18.26 \\
ea6f316259 & H8SeS3 & P2\_1/m & 130.0 & 110.0 & 105.0 & 113.0 & 118.11 $\pm$ 12.84 & 109.68 $\pm$ 12.99 \\
ec9086ecaf & Zr7ScH30 & R3 & 115.0 & 114.0 & 155.0 & 145.0 & 30.14 $\pm$ 4.48 & 40.25 $\pm$ 30.17 \\
ecddef9b73 & YMg5(TcH7)2 & P-6m2 & 32.0 & 12.5 & 34.05 & 47.0 & 49.31 $\pm$ 6.99 & 21.82 $\pm$ 13.99 \\
eda2f14253 & PH3S & P2\_1/c & 115.0 & 107.0 & 100.0 & 106.0 & 103.12 $\pm$ 12.5 & 125.14 $\pm$ 26.17 \\
f10f4dcd72 & Th3B2CN3 & Cm & 11.0 & 13.0 & 25.0 & 14.0 & 17.21 $\pm$ 4.26 & 15.19 $\pm$ 10.6 \\
f2e3fa7596 & BaNbH5 & Pnma & 12.0 & 14.0 & 10.0 & 14.0 & 43.76 $\pm$ 11.07 & 18.28 $\pm$ 15.65 \\
fdd1c102aa & Li8TaNbN6 & I2\_12\_12\_1 & 11.5 & 39.0 & 13.0 & 16.1 & 19.29 $\pm$ 2.27 & 18.32 $\pm$ 15.5 \\
ff16239bad & B13CN & R3m & 15.0 & 13.5 & 10.0 & 17.0 & 22.31 $\pm$ 5.4 & 32.45 $\pm$ 17.85 \\
ff75fdc3e2 & LiMg2AlB28 & C2/m & 10.0 & 29.0 & 20.05 & 25.0 & 20.59 $\pm$ 4.42 & 10.64 $\pm$ 6.61 \\


\end{longtable}

\twocolumngrid    

\bibliographystyle{apsrev4-2}
\bibliography{apssamp}

\clearpage

\end{document}